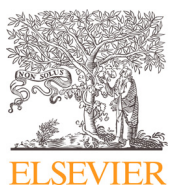
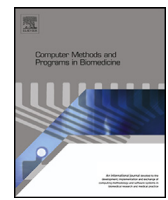

# Robust heartbeat detection using multimodal recordings and ECG quality assessment with signal amplitudes dispersion

Zahra Rezaei Khavas, Babak Mohammadzadeh Asl*

*Electrical and Computer Engineering Department, Tarbiat Modares University, Tehran, Iran*

**a b s t r a c t**

*Background and Objectives:* The electrocardiogram (ECG) is a bioelectric signal which represents heart's electrical activity graphically. This bioelectric signal is subject of lots of researches and so many algorithms are designed for extracting lots of clinically important parameters from it. Most of these parameters can be measured by detecting R peak of the QRS complex in ECG signal, but when ECG signal is corrupted by different kinds of noise and artifacts, such as electromyogram (EMG) from muscles, power line interference, motion artifacts and changes in electrode-skin interface, detection of R peaks becomes hard or impossible for algorithms which are designed for heart beat detection on ECG signal. In modern patient monitoring devices often not only one ECG signal is recorded but also so many other biological signals are simultaneously recorded from the patient which some of them, such as blood pressure (BP), are containing useful information about the heart activity which could be very helpful in making the heart beat detection more robust.

*Methods:* In this study, a new method is introduced for distinguishing noise free segments of ECG from noisy segments that uses samples amplitudes dispersion with an adaptive threshold for variance of samples amplitude and a method which uses compatibility of detected beats in ECG and some of other signals which are related to the heart activity such as BP, arterial pressure (ART) and pulmonary artery pressure (PAP). A prioritization is applied in other pulsatile signals based on the amplitude and clarity of peaks on them, and a fusion strategy is employed for segments on which ECG is noisy and other available signals in the data, which contain peaks corresponding to R peak of the ECG, are scored in a three steps scoring function.

*Results:* The final scores achieved by the proposed algorithm in terms of average sensitivity, positive predictive value, accuracy and F1 measure on the database which is freely available in Physionet Computing in Cardiology Challenge 2014 are respectively 95.47%, 96.03%, 93.11% and 95.62%.

*Conclusions:* The results show the outperformance of the proposed method against other recently published works.

## 1. Introduction

Cardiovascular diseases are one of the main causes of death; a great percent of all sudden and unexpected global deaths was made up by the people who died because of cardiovascular diseases [1]. Electrocardiogram)ECG(signal is a tool to represent the electrical activity of heart graphically; an ordinary heart beat in the ECG signal is containing P-wave, QRS complex and T-wave and it can be easily measured by two or more electrodes on surface of the patient's body [2]. One of the main uses of ECG signal is for detection of heart beats. Detection of heart beats is mainly studied for estimation of heart rate (HR) and heart rate variability (HRV) which are clinically important parameters [3], these estimations are usually accomplished by the detection of the QRS-complexes and calculation of RR intervals between each two consecutive cardiac contractions. Using ECG for heart beat detection is the best choice because this biological signal's acquisition process is simple and noninvasive, thus this signal is most used in estimation of HR and heart beats location rather than other biological signals which contain noticeable signs of heart beats such as blood pressure (BP), pulmonary artery pressure (PAP), or photoplethysmography (PPG). So, most of the heart beat detection algorithms designed yet are based on ECG and indeed they are R peak detectors [4], a complete review of more recent ECG based peak detectors can be found in [5]. Also, there are some algorithms that measure HR using other biological signals but ECG such as [6] and [7] which respectively estimate HR based on mobile data which uses a stream of picture frames for heart beat information extraction and plethysmography which measures heart rate based on oxygen saturation of blood, optically.

* Corresponding author.
  *E-mail address:* babakmasl@modares.ac.ir (B.M. Asl).

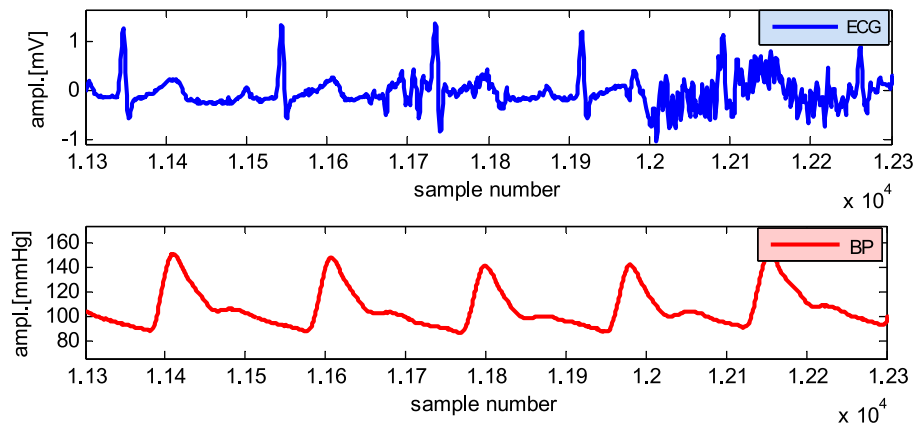

**Fig. 1.** A noisy ECG segment and corresponding BP segment, label of Y axis (ampl) represents the amplitude.

Large quantities of data are acquired from a patient in different monitoring devices connected to the patient in intensive care unit (ICU) by connecting noninvasive and invasive sensors. So, several multimodal databases are acquired of such monitoring devices and these databases are used for so many investigations [8,9]. This acquired data might include so many different biological signals such as ECG, PPG, electroencephalogram (EEG) and electromyogram (EMG).

In addition to ECG, some of other signals that are existing in the data are related to the heart activity and in some ways displaying heart beats, such as BP or stork volume (SV), (e.g. as soon as the heart contraction occurs, QRS complex in the ECG appears, as well as a peak in SV, PPG, BP and other signals which are related to heart activity (as shown in Fig. 1)). As stated by the Physionet Challenge 2014: robust detection of heart beats in multimodal data, other signals could be used to improve heart beat detection robustness [10]. Generally, the main aim of this challenge is to motivate the development of algorithms for robust heart beats location estimation in long-term multimodal recordings.

The main goal of this study is fusing the information about heart beats acquired of all these signals to make a more accurate decision about heart beats location and a more robust HR estimation. This might also be used for false alarms reduction in ICU which was the subject of Physionet Computing in Cardiology 2015 [11].

There are some studies about detection of heart beats on multimodal data. A comparison of beats regularity among ECG and BP in 5 s window, detection of noise free periods by calculation of PSD and replacement of BP beats in noisy ECG segments, beat selection among ECG and BP by calculation of quality index for ECG and BP signals in 10 s windows, and a calculation of BP beats delay compared to ECG beats and detection of beats in noisy segments based of the delay were reported in [4,12,13], and [14], respectively. In [15], a robust algorithm for heart beat detection on multimodal data, based on slope and peak-sensitive band-pass filters, was presented. In [16], an algorithm was presented which uses ECG, BP, SV, PPG, EEG, and electrooculogram (EOG) signals for heart beat detection and fuses the extracted information from these signals with a voting approach. [17] Presented an algorithm that uses probabilistic model for detection of heart beats from ECG and BP based on hidden semi-Markov model. The method presented in [18] uses ECG, BP, EEG, EOG and EMG for multimodal heart beat detection by employing a quality assessment strategy for deciding which signal gives the best results in each segment.

In this paper, a new method for estimating the accurate location of heart beats in multimodal recordings is presented; in addition to ECG, other signals such as BP, arterial blood pressure (ART), PAP and SV are used in this method for detection of the heart beats location in a robust way. Reading the labels of the data channels is done automatically with a code available in the WFDB toolbox [19]. Several peak detectors are employed for initial heart beat detection in this method. In the first phase of peak location estimation, this algorithm mainly uses ECG, BP and ART signals and introduces a new method that uses compatibility of detected beats in these signals for assessing the quality of these signals. Also, this paper introduces a unique method for assessing quality of ECG signal by using dispersion of samples amplitude. Our proposed method employs a fusion strategy that contains scoring all sets of detections on different signals by different detectors with three scoring criteria.

In most of the methods mentioned here for heart beat detection on multimodal data, only two signals were used and only one detector was applied to each of the signals, all of these methods used only one quality assessment criteria for selecting the best set of beats in each window which is only based on beats comparison on two signals or regularity of beats on each signal, and after quality assessment one signal was completely removed and beats of the other signal were selected as final beats in a segment. The presented method of this paper uses more than two signals of each record for heart beat detection, on each of the signals more than two beat detectors are used for heart beat detection, quality assessment of the signals takes place in 3 different steps and a new quality assessment criteria for ECG is presented in this method which is adaptive and it is based on samples amplitude dispersion; after the quality assessment one of the signals is removed only if its quality is under the specified threshold, otherwise no signal is removed and all of the signals are sent to the final scoring step.

## 2. Data

### 2.1. Data description

The database used in this study is acquired from Physionet/Computing in Cardiology Challenge 2014 [20]. This database contains 200 records which are taken from 2 different databases, challenge 2014 training set and challenge 2014 extended training set. Each record in these two databases belongs to a human adult (men or women) with a wide range of different cardiac diseases. The records of these two databases mostly have 10 min length but in some records the length of the recorded data is less than 10 min, the minimum length of the records in these two databases is 19.9 s, the average length of all the records in these two database is 563.1 s and the standard deviation of the records length in these two databases is 118.2 s. Each of these records is composed of 4 to 8 different signals which are recorded simultaneously, such as ECG that represents the electrical activity of heart graphically, BP that represents the pressure that blood carries to the special vessel's wall in blood circulation, ART that refers to the pressure in large arteries of the systemic circulation, pulmonary artery pressure (PAP) that represents the pressure inside the vessel that carries blood from heart to the lungs (pulmonary artery), central venous pressure (CVP) which is the pressure of blood in the vena cava, close to the right atrium of the heart, EEG that represents the brain's electrical activity, EOG which shows the electrical activity of eyes, EMG which records the electrical activity of muscles, SV that shows the volume of blood pumped out of heart in each contraction as a signal, oxygen level ($SO_2$) that shows the amount of oxygen-saturated hemoglobin in the blood, respiration (RESP) which shows the movements of oxygen and other gases available in the air into and out of the lungs, and carbon dioxide level ($CO_2$) that shows the carbon dioxide concentration in blood. Some of these signals are dependent and some others are independent from activity of heart.

**Table 1**
Details of databases used in this study.

| Database name | Number of records | Records length (Mean ± std in seconds) | Noise level | Sampling frequency | Mostly containing signals | Number of signals in the records |
|---|---|---|---|---|---|---|
| Challenge 2014 training set | 100 records | Mostly 10 min (some records are shorter) (563.1 ± 118.2) | Low noise level | 250 Hz for all records | ECG, BP, SV, EEG and EMG | 4 to 8 |
| Challenge 2014 extended training set | 100 records | Mostly 10 min (some records are shorter) (521.7 ± 160.6) | High noise level in some records | varies between 120 to 1000 Hz for different records | ECG, ART, PAP, CVP and respiration | 4–8 |

There is no regulation or rule in the arrangement of signals in each record and it is completely different for each record, the only fixed thing about all records is that the first signal is always ECG (in different records, ECG signals from different leads are available; in few records, more than one ECG are available). Challenge 2014 training set contains 100 multimodal records with the same sampling frequency of 250 Hz, number of noisy signals in this database is a lot less than noisy signals of challenge 2014 extended training set (this issue is examined by applying different peak detectors on the two databases separately and comparing the results), records of this database mainly contain BP, SV, EEG and EMG signals in addition to ECG signal. Number of the multimodal records in the challenge 2014 extended training set is equal to the challenge 2014 training set but the sampling frequency of the records in this database is different for each recording, and it varies between 120 to 1000 Hz for different records, but the sampling frequency is same and fixed for all the signals in one record. Records of this database mainly contain ART, PAP and CVP signals in addition to ECG signal. This database contains records from patients that have cardiac pacemaker. In order to score the algorithm and evaluate validity of detected beats, a reference annotation set is provided by some experts for each recording and a total of 151,032 beats were annotated in these two databases. You can see all the details explained in this section about these two databases in Table 1.

### 2.2. Data selection

As mentioned in previous section, many different signals available in each record. Fig. 2 shows an example of a multimodal recording segment. We separate these signals into two groups: the first group comprises signals that are directly associated with the activity of heart in some ways, such as ECG, BP, ART, PAP, CVP and SV; The second group contains signals that are not associated with activity of heart, such as EEG, EMG, EOG, RESP, $SO_2$ and $CO_2$. Note that ECG signal has an electrical origin and in a same manner the origins of EEG, EOG and EMG are the electrical activity of brain, eyes and muscles respectively and they are recorded with skin connected measurements. As these signals have lower amplitude in comparison with ECG signal, sometimes effects of electrical activity of heart could be seen on these signals and it is called ECG interference, so sometimes heart beats could be seen on these signals. These signals could also be used for heart beat detection as they are used in [18] but in this work we have not used them.

## 3. Methods

### 3.1. Algorithm overview

Main goal of the presented method is designing an algorithm for robust heart beat detection that uses ECG and other biological signals which are related to the heart activity and the effects of heart activity can be clearly observed on them. This algorithm consists of multiple stages as shown in the block diagram of Fig. 3.

First stage of the algorithm is signal selection. In this stage, the signals which are going to be used for heart beat detection in this study (ECG, BP, ART, PAP, CVP and SV) are selected from each record. After selecting the signals, next stage is denoising. In this stage, the signals are denoised by the use of wavelet decomposition. First of all, signals are decomposed by wavelet decomposition to different frequency bands and after that frequency bands which contain more noise effects are removed and hence, the remaining signal contains less noise effects. In fact, wavelet decomposition of signal and removal of some frequency bands work as a bandpass filter. Denoising strategy in this study is more complete and complex for ECG signal, but it is simple and fast for other signals. Denoised ECG is then normalized and optimized for better peak detection. After denoising, there is a peak detection stage. In peak detection stage, several peak detectors applied to each of the signals which have been chosen in signal selection phase for heart beat detection such as wabp [21] and gqrs from WFDB toolbox [19], a detector which is designed based on Pan and Tompkins [22], a general detector which is designed for this work named Window (Absolute maximum), a general detector named adaptive threshold [4] and another general detector which is based on MATLAB's find peak command [23]. All of the detections of different peak detectors on any signal are stored to be used in the fusion stage if it is needed. In the next stage, named delay correction, lag times among beats detected on all the signals compared to beats detected on ECG signal are measured and these lag times are compensated by shifting detected beats of other signals but ECG, as much as measured delay. After that, there is a quality assessment stage on which the algorithm checks if ECG is clean enough for heart beat detection or it is needed that a fusion of detected beats in all signals be used for heart beat detection. As shown in Fig. 4, in this stage, the algorithm checks each 5 s window of the signal, separately. Quality assessment stage consists of three different steps. In the first step, quality of ECG and BP (or ART) signals are checked for whole signal length by checking compatibility of all the detected beats on these two signals. In the second step, quality assessment is done by beat comparison in 5 s windows and quality of each window is checked, and in the last step, which only runs when incompatibility occurs in second step, quality of ECG signal is checked by calculation of sample dispersion.

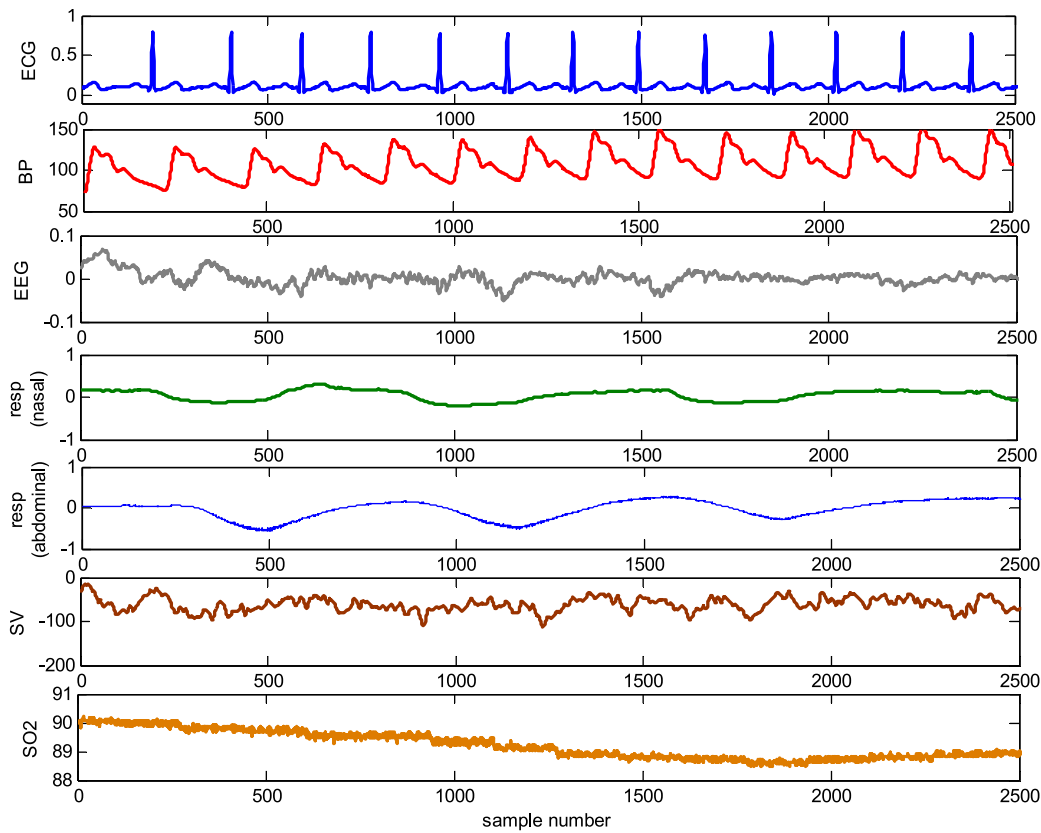

**Fig. 2.** Example of a multimodal recording segment. In this record, the BP and SV signals can be used in order to improve the performance of the heart beat detection in situation that the ECG would have been corrupted by noise. The sampling frequency of the signal in this record is 250 Hz.

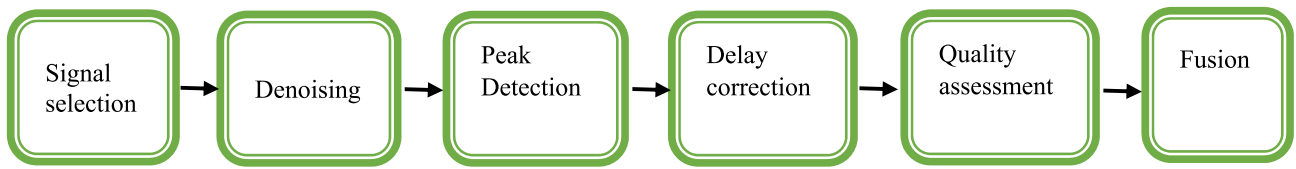

**Fig. 3.** Block diagram of the proposed method.

Finally, the fusion stage is employed for the segments of ECG signal with low signal quality estimated by quality assessment stage. As shown in Fig. 5, in the fusion stage, three different scoring criteria are employed based on regularity of beats, physiological range of beats rate and compatibility of beats detected by all detectors on all of the candidate signals. This stage is designed to score all of the detections of all detectors on all signals on the specified segment and select the best set of detections on that segment. For checking if any surplus or missed beat occurred or not, a search back stage is designed at the end of the algorithm.

### 3.2. Preprocessing

Presented method in this paper uses so many peak detectors for heart beat detection from different signals that some of them contain an initial preprocessing stage in them such as gqrs, wabp [21] and Pan and Tompkins [22]. The gqrs algorithm is unpublished but in some references such as [14] it is mentioned that this algorithm contains denoising strategies; wabp algorithm contains a low pass filter to suppress high frequency noise that might affect the BP onset detection and a slope sum function to enhance the upslope of the BP pulse and to suppress the remainder of the pressure waveform [21]; Pan and Tompkins algorithm contains a band pass filter (which consists of cascaded low pass and high pass filters) for attenuating noise effects [22]. So, these detectors remove noise and artifacts of signals automatically and hence, input signals for these detectors do not need any preprocessing. For those other detectors which do not contain any preprocessing stage, a simple preprocessing method is designed in this work. Preprocessing method designed in this work contains a more complete and complex process on ECG, but it is sufficed to a simple and fast preprocessing on other signals, because these signals are not going to be used when they are noisy and they are used whenever they have high quality and the location of heart beats are clear on them.

There are many studies about denoising ECG with wavelet [24–26] which are mostly based on creating a bandpass filtering strategy by the use of wavelet transform. In this study, we used a method for denoising ECG by the use of wavelet that is intuitive, this method consists of decomposition of signals to different frequency bands by wavelet decomposition and using the wavelet coefficients of frequency levels that the effects of QRS complexes are clearly evident on them for creating a new signal. As you can see in Fig. 6, when ECG signal is decomposed to different frequency levels by the use of wavelet decomposition, effects of the QRS complexes are evident clearly on decomposed levels d2, d3 and d4. So, new signal is generated with the summation of wavelet coefficients of those levels which QRS complexes effects are clearer on them. After that the new signal is normalized and optimized with the use of Eqs. (1) and (2), respectively.

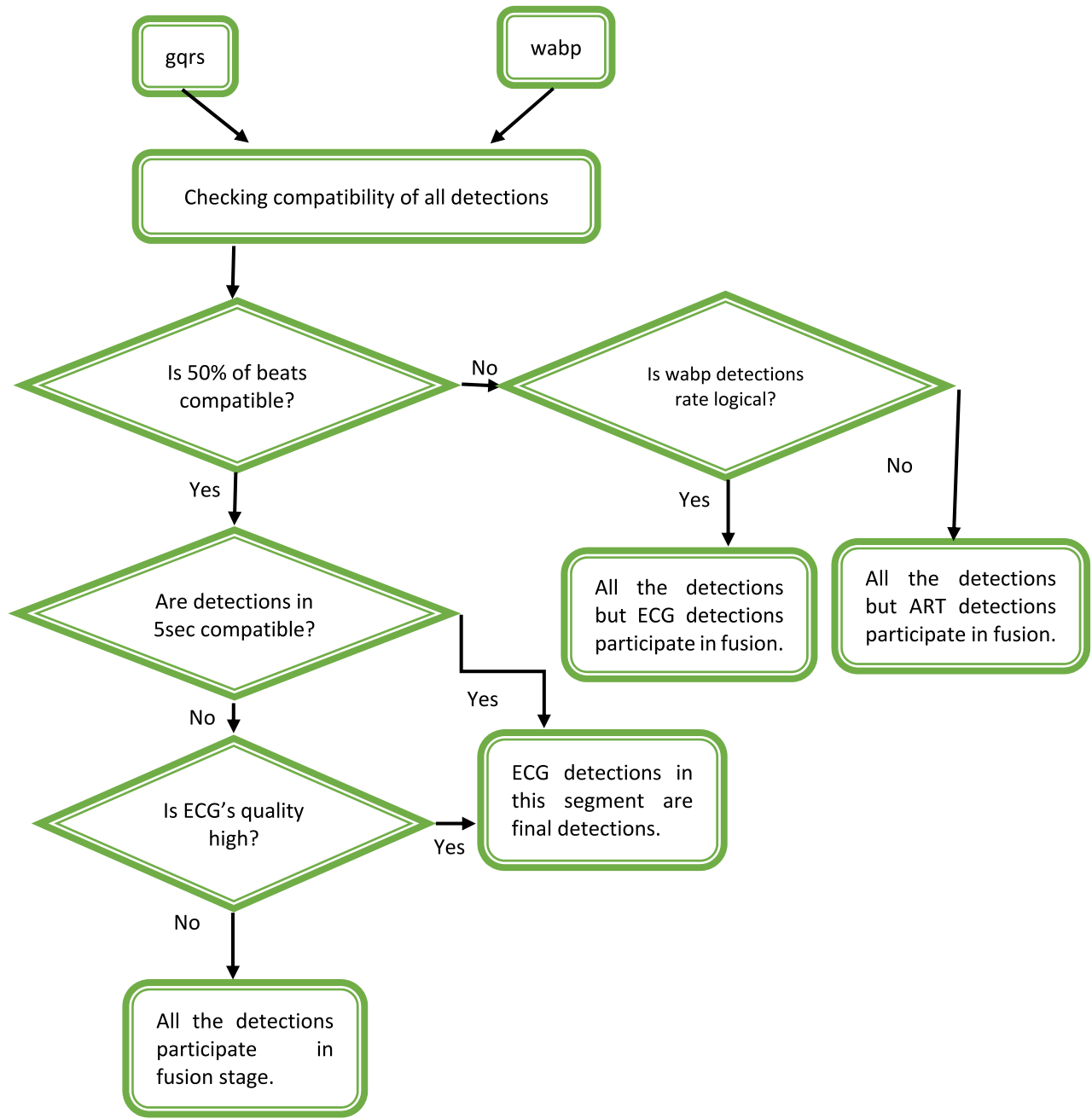

Fig. 4. Block diagram of the quality assessment stage of the proposed method.

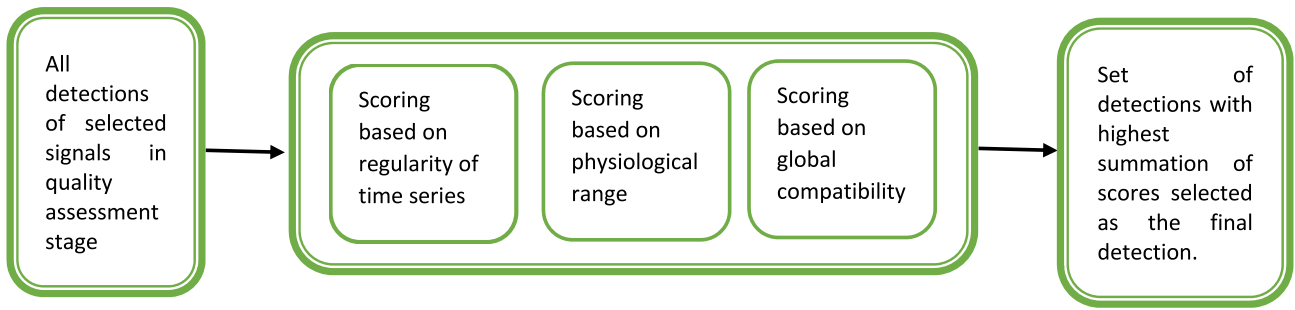

Fig. 5. Block diagram of the fusion stage of the proposed method.

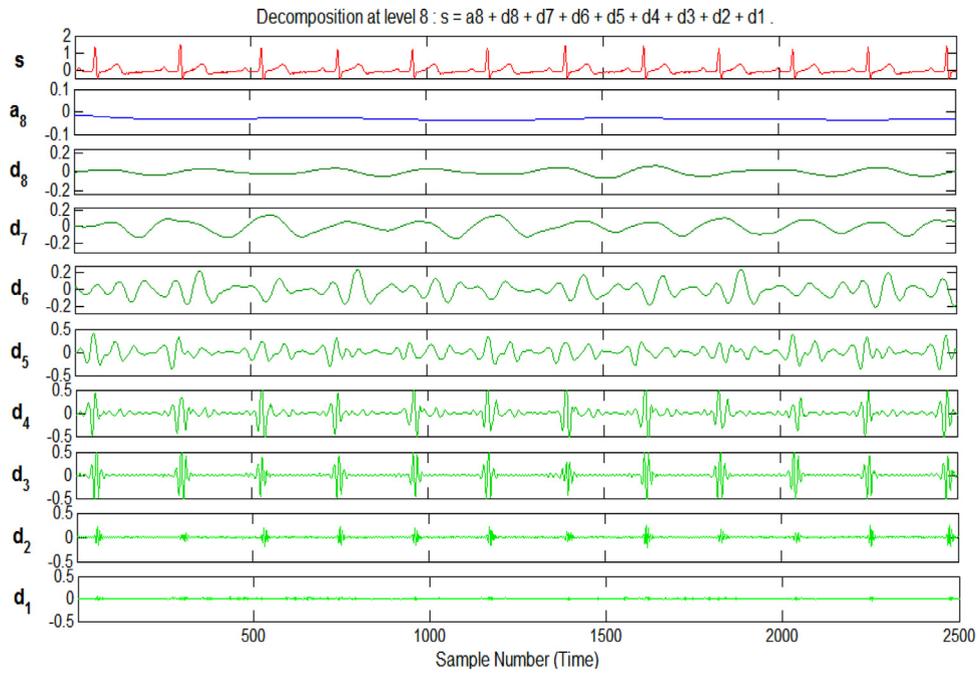

**Fig. 6.** Wavelet decomposition of ECG signal with mother wavelet db 6; (s) Raw ECG signal, in this figure, Y label shows the amplitude of signal samples in millivolt and X label shows the sample number of ECG or time steps. d1, d2, d3, d4, d5, d6, d7, d8 and a8 are wavelet coefficients of ECG signal decomposition, in these figures, Y labels show amplitude of wavelet coefficients (scale) and X labels shows the coefficients index.

$$\text{Normalized ECG} = ECG_1 = (ECG - \text{mean}(ECG)) / \max(ECG) \quad (1)$$

$$\text{Optimized ECG} = ECG_2 = (4 \times ECG_1)^2 \quad (2)$$

Preprocessing stage for all other signals but ECG is just designed in such way that removes baseline drift, because these signals are used only in the situation that they are not noisy. Removing baseline drift from these signals is performed by decomposing these signals to different frequency bands by the use of discrete wavelet decomposition (haar mother wavelet) and omitting wavelet coefficients of decomposed level d1, which baseline drift's effects could be seen on it.

### 3.3. Beat detection

As described in Section 2.2, several signals are chosen from each record for heart beat detection; in all the records at least one lead of ECG signal is always available, if any of BP, ART, PAP and SV signals exists in a record it could be considered as a candidate for heart beat detection. It is decided to apply more than one detector to each candidate signal. As there are several candidate signals for heart beat detection from each record, so too many detectors are needed in this algorithm. In order to reduce the running time of the algorithm, three simple and fast general peak detectors, which could be applied to any pulsatile signal, are employed instead of using different peak detectors. These three detectors are applied to all of the candidate signals. In addition to these three general peak detectors, two proprietary beat detectors for ECG and one for BP are used (if there was no BP available in a record, this detector works on one of the ART or PAP signals if any of them is available, and if both of them be available on a recording, priority is with the ART signal). So, in any recording after signal selection, 5 peak detectors are applied to ECG signal, 4 peak detectors are applied to the pressure signal with the lowest index and three peak detectors are applied to other candidate signals. Using this proprietary detectors has an advantage; these algorithms have more complex and advanced processing levels and hence more accurate results of beat locations for ECG and the pressure signal with the lowest index could be acquired by them, which is needed for quality assessment stage, more descriptions are given in Section 3.5. The difference between numbers of detectors for different signals provides some kinds of weighting to signals, more descriptions are given in Section 3.6.

#### 3.3.1. General peak detectors

**Window (Absolute maximum):** One simple way for detection of peaks in pulsatile signals is to employ window based strategies and finding the point which has the maximum amplitude. In this approach, window length is considered to be 0.8 s for all the signals (0.8 is calculated experimentally). As the window length is fix, it cannot handle the need of changes in window length for HR variability, then when changes occur in HR it causes false positives (false positive which is widely known as FP, denotes the number of falsely detected beats) and false negatives (false negative which is widely known as FN denotes the number of undetected beats) in the detections of this detector. In order to solve the mentioned problem, FP remover and FN remover functions are designed to peruse interval of consecutive beats. These two functions compute the min and max permissive intervals of two consecutive beats dynamically. If interval of two consecutive beats exceeds the max, algorithm looks for FNs and if this distance is lesser the min, algorithm looks for FPs. The min and max permissive intervals are calculated by Eqs. (3) and (4):

$$\text{min permissive interval} = 0.5 \times (\text{mean of last 10 RR intervals}) \quad (3)$$

max permissive interval = 1.3 × (mean of last 10 RR intervals). (4)

Whenever the interval between two consecutive beats is lesser than the min permissive interval, the algorithm considers that a FP detection occurred. After that the RR intervals of these two beats with the last beat before these two beats are calculated and the one which the value of its RR interval with the last beat before these two beats is closer to the mean of last 10 RR intervals is accepted as a TP and the other detection is considered as a FP and will be removed. Whenever the interval between two consecutive beats is greater than the max permissive interval, the algorithm considers that a FN detection occurred. Then the algorithm looks for the local maximum among these two beats and considers this local maximum as a correct beat.

**Adaptive threshold:** Another simple way for detection of peaks in any pulsatile signal is using a threshold based strategy; any point that has an amplitude which exceeds a specified predetermined threshold is known as a peak. The threshold in this approach is calculated as follows:

Threshold = 4 × (mean(abs(data))). (5)

It means that the threshold is four times greater than the mean of absolute value of all the signal samples [4]. This detector has the problem of occurrence of lots of FPs and FNs same as the window detector (e.g. amplitude of a T-sharp might be high enough to be considered as a peak, or in noisy segments of signal, lots of noise peaks appear in a short segment of signal which all of them might be detected by the detector, or if a real peak has low amplitude it might be missed). In order to solve the mentioned problem, FP remover and FN remover functions are used on the detections of this detector too.

**Local maximum (Find peaks):** This general peak detector is based on the MATLAB's findpeaks command [23]. This command has different peak detection strategies and you can choose the most suitable ones for your job. The findpeaks approach used in this paper is based on minimum distance of two candidate peaks. Using this command needs an input parameter, this parameter is minimum distance possible of two consecutive beats. In this study, the minimum distance is calculated by use of detected beats via window and threshold detectors as follows:

Min_distance = 0.8 × mean ([diff(peaks_AbsoluteMAX)
diff (peaks_threshold)]) (6)

where calculated Min_distance in Eq. (6) is used as input parameter for local maximum beat detector, diff and mean are respectively difference of detected beats and average of these differences.

#### 3.3.2. ECG proprietary detectors
- **gqrs:** This peak detector is suggested by Physionet to be used for peak detection in ECG signal; designer of this peak detector is George Moody, and this detector is available on Physionet (freely) but the algorithm is not published.
- **Pan & Tompkins:** This algorithm is very famous for heart beat detection in ECG signals and contains different approaches for the initialization of two sets of thresholds that each of them contains one upper and one lower thresholds. This algorithm doesn't have any search back strategy, which makes it suitable for an online heart beat detection and so this algorithm needs minimal memory [22].

#### 3.3.3. BP proprietary detector
- **wabp:** This detector is an open source algorithm that works based on length transform [21]. This detector detects onset of peaks in hemodynamic signals such as BP, ART and PAP but this algorithm could not use CVP signal for peak detection. This peak detector also could be applied to signals named pressure or general pressure. If a record contains more than one of the named signals, wabp peak detector uses signal with the lowest index. This algorithm is available on Physionet [27].

### 3.4. Delay correction

Signals chosen for this study (signals from the first group) were divided into two clusters according to their origin. The first cluster entails those signals which are originated from electrical activity of heart and propagate almost instantaneously. This cluster just contains ECG signal. The second cluster consists of those signals which are correlated with heart's mechanical activity. Heart beats detected in this cluster reflect the change in blood pressure pulse waves moving along the arteries as a consequence of the heart's contraction that occurs in systole, such as BP, ART, CVP and PAP. SV signal shows the volume of blood ejected by the left ventricle in each contraction, so SV signal could also be classified in the second cluster of signals. Signals of the second cluster have a delay in comparison to signals of the first cluster so beats available in the second cluster appear with a delay compared to beats available in ECG (first cluster signals). Mechanical activity of heart needs up to hundreds of milliseconds to travel to the site that the sensor is connected.

Delay time for each signal compared to ECG is different from other signals, in addition to that, delay time for each beat in one signal varies from the delay time for other beats and it is not a constant value for each signal. Note that in addition to the variation of delay times between different beats of one signal, delay times for different signals, depending on the record site or location of the sensors and the technologies employed for recording, are different (e.g. for the pressure sensors that are located centrally near by the heart location, delay time might be on the order of tens of milliseconds but for the sensors that record pressure on fingers, this delay might be above 400 ms).

We considered a constant delay time for each signal, because the variance of these delay times between different beats in one signal is negligible in this study compared to the allowed detection neighborhood of 150 ms by Physionet/Computing in Cardiology Challenge 2014 [20], so the mean value of them could be used as the delay time for all beats in one signal. Mean delay time for BP signal compared to ECG is estimated 200 ms [13].

### 3.5. Quality assessment

Quality assessment stage contains three steps, global quality assessment, local quality assessment and ECG quality assessment. Both of local and global quality assessment steps are used to investigate quality of main cluster's signals and use only this cluster's signals. In global quality assessment step, detections of main cluster's signals are compared for whole record length and if more than 50% compatibility occurs, local quality assessment runs. ECG quality assessment step only runs when local quality assessment step shows an incompatibility in the detected beats of main cluster's signals. These three steps are shown in the Fig. 4, and explained in more details below.

- **Global quality assessment step**: In this step, quality of main cluster's signals are controlled by comparing detections of 'gqrs' on ECG and detections of 'wabp' on lowest index of BP, ART and PAP in the record, as wabp detector chooses the pressure signal with lowest index automatically. If more than 50% of detected beats of these two detectors are compatible and for each gqrs detection a corresponding beat on wabp detections be available in 100 millisecond neighborhood, (in more than 90% of records, expected compatibility occurs), second and third quality assessment steps run; otherwise only one of the main cluster's signals is selected to be fused with the signals of alternative cluster.

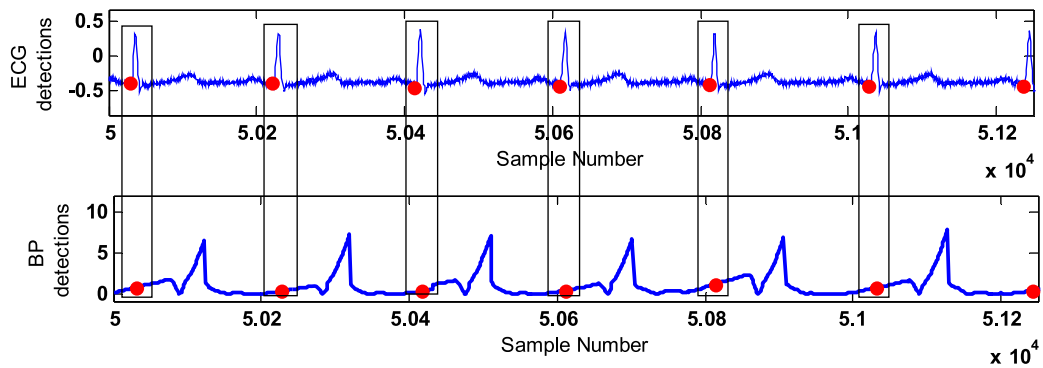

**Fig. 7.** An ECG and BP corresponding segments on which detected beats are completely compatible.

This step is designed in order to identify if any of main cluster's signals does not have satisfying quality for whole signal length, and then remove it. After checking the compatibility, if less than 50% of beats are compatible, one of the two signals of main cluster that the detections ratio of that is not in the logical physiological range, is removed. For example, flat signals which are not correctly recorded and there is no beat available on them for whole record length and noisy signals which are oscillating for whole signal length and lots of incorrect beats are detected on them by peak detectors are identified and removed in this step. This step also identifies ECG signals which contain paced beats (ECG signals which are recorded from people that have pacemaker on their heart) and omits them. These signals traditionally were omitted from databases that were used for evaluating ECG beat detectors, but in this study because beat detection is not based on only one ECG signal, even if ECG signal is omitted, detection of heart beats is done by use of other signals.

- **Local quality assessment step:** When quality of signals for whole signal length was checked and confirmed, quality of ECG in short segments must be checked to identify segments in which detected beats on ECG are not reliable and must be replaced by detected beats of other signals. In order to fulfill this aim, compatibility of candidate peaks of main cluster's signals is checked by comparing detections of 'gqrs' on ECG and detections of 'wabp' on lowest index of BP, ART and PAP in the record, in 5 s windows. If detected beats of ECG be completely compatible with detected beats of other signal of main cluster (for each ECG beat a corresponding beat on other signal be available in 100 ms neighborhood), ECG detections on that window considered as final detections and the algorithm do not run the ECG quality assessment step and do not enter the fusion phase, otherwise in condition that even one mismatch occurs, quality of ECG must be checked in the third quality assessment step. In Figs. 7 and 8 respectively a compatible and an incompatible segments of signals in a 5 s windows are shown.
- **ECG quality assessment step:** In this step, algorithm wants to check if the incompatibility occurred in last step is as a result of noise on ECG or not. If ECG signal is considered to be noisy in this step, this segment of ECG do not take part in fusion and fusion is done among other signals and if ECG segment is considered to be clean, detected beats on ECG in that segment are considered as the final detections on that segment. ECG signal in noisy segments is very unstable and lots of peaks and valleys occur in a very short segment of the signal, which cause significant increase in dispersion of the samples amplitude. This feature of ECG signal could be used for identifying if a window of signal is clean or not. Dispersion of samples amplitude could be measured by calculating the variance of the samples amplitude. In order to identify noisy windows from clean windows, wavelet coefficients of ECG signal decomposition are used instead, because the difference in the variances of noisy and clean windows is more specific in wavelet coefficients of level d2 when ECG signal is decomposed to different frequency bands by the use of discrete wavelet transform. Note that db6 mother wavelet is used for ECG decomposition. In Fig. 9, wavelet coefficients of d2 level of a clean ECG window and a noisy ECG window are shown and variance of the wavelet coefficients of them are reported.

Two thresholds are needed, clean segments variance threshold and noisy segments variance threshold which are calculated for each segment adaptively. Clean segments variance threshold is considered as mean value of compatible ECG segments variance plus standard deviation of them and noisy segments variance threshold is considered as mean value of incompatible segments variance minus standard deviation of them.

Incompatible 5 s ECG windows which variance of their wavelet coefficients is less than clean segments variance threshold are considered as clean segments and no other processing is needed in these segments and detected beats in these ECG segments are considered as final beats. For segments which variance of their wavelet coefficients is among these two thresholds, all beats detected in these segments from all signals are scored in the scoring step and final beats are identified through scoring step. For segments which variance of their wavelet coefficients is greater than noisy segments variance threshold, ECG is removed from the signals and final beats are identified through scoring step between all detections but ECG detections.

### 3.6. Scoring

In this step, each set of detected beats is scored by three different scoring standards which are listed below:

I. **Regularity of time series**: In noisy segments of ECG signals, occurrence of FPs and FNs in the detections of all detectors makes consecutive RR intervals unsteady, so regularity of RR intervals could be a good feature to investigate if a segment is noisy or not. Hence, the variance of RR intervals for each set of detections is calculated and the set of detections corresponding to the time series with the smallest variance of RR intervals achieves the highest score which is equal to n, where n is number of all sets of detections in that segment. The score for other sets of detections is equal to n-k, where k is that set's ranking when we sort the sets with respect to their variance in ascending order. Fig. 10 shows two ECG segments, one of them is clean and the other is very noisy; the initial detections in them as candidate heart beats are shown in both segments and distance between each two consecutive detections (RR interval) is shown by two direction arrows. It is clear that in clean segment, detections are more regular and so variance of RR intervals is smaller than noisy segments. Note that if a set of detections contained less than two beats within the 5 s window, that set was considered as the last set in ascending order.

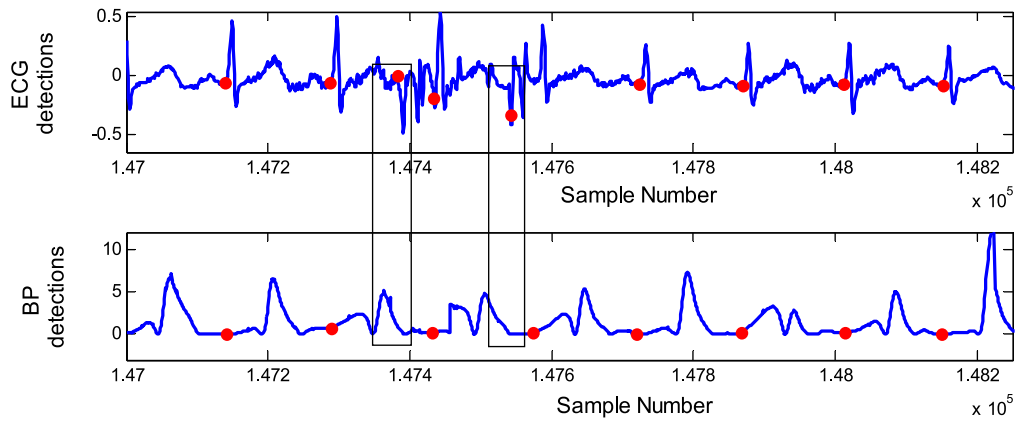

**Fig. 8.** An ECG and BP corresponding segments on which detected beats are incompatible.

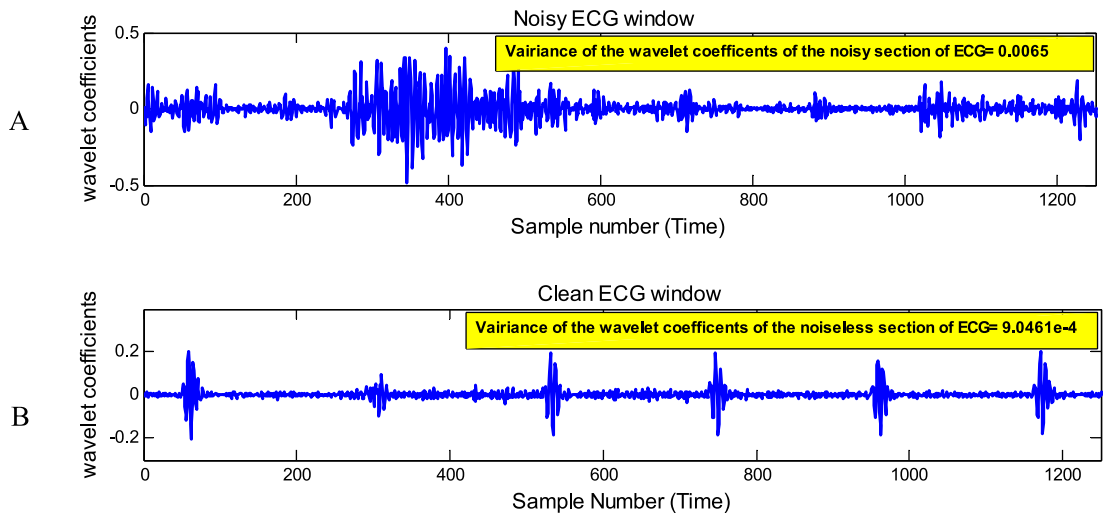

**Fig. 9.** Wavelet coefficients of d2 level of two ECG segments with same length (5 s) and different noise levels and their variances of wavelet coefficients, (A) is more corrupted by noise than (B) and its variance of wavelet coefficients is about 7 times bigger than (B).

II. **Physiological range**: In noisy segments of ECG signals, occurrence of lots of peaks and valleys deforms the signal and cause lots of FPs in detected beats and candidate detections ratio, which is the number of detected beats per second, increases suddenly. In the segments the signals of which are not recorded correctly, number of detected beats per second decreases suddenly. So, checking the physiological range of detected beats in each segment could be a good criterion for investigating if detected beats in a segment of signal are reliable or not. Hence, detection ratio for each set of detections is calculated and the algorithm checks if the detection ratio of a set is in the range of 0.8–1.9 [4]. Sets of detections whose detection ratio were in the range of 0.8–1.9 (this range is the logical physiological range for heart beats) achieve highest score which is equal to $n$ ($n$ is the number of sets of detections). If detection ratio of a set of detections is not in the 0.8–1.9 range, algorithm checks if its detection ratio is in the range of 0.5–2.5. Sets of detections whose detection ratio were in the range of 0.5–2.5 achieve the score equal to $\frac{2}{3} n$ and other sets of detections achieve the score of $\frac{1}{3} n$.

III. **Global compatibility**: If a real heart contraction occurs its effects first of all appear in ECG and after that in all signals which are related to the heart's activity, so whatever the number of unanimous signals for a detected beat is greater the more reliable that beat is. It could be a good standard for scoring detected beats in a segment. For this purpose, global compatibility assessment for each beat in each set of detections is done separately; score of each beat is equal to the number of compatible beats with intended beat in all sets of detections. Score of each set is equal to summation of scores achieved by all its beats. The set with the highest summation achieves the highest score which is equal to n, where n is number of all sets of detections in that segment. Score of other sets is equal to $n-k$, where $k$ is that set's ranking when we sort the sets with respect to their summation in ascending order.

Finally, results of all these three scoring steps are assembled to specify the best set of detections in that segment.

### 3.7. Search back

As the final set of heart beat detections in this strategy is assembled from detections in 5 s windows of different signals, in final phase of this method a search back strategy is designed in or-

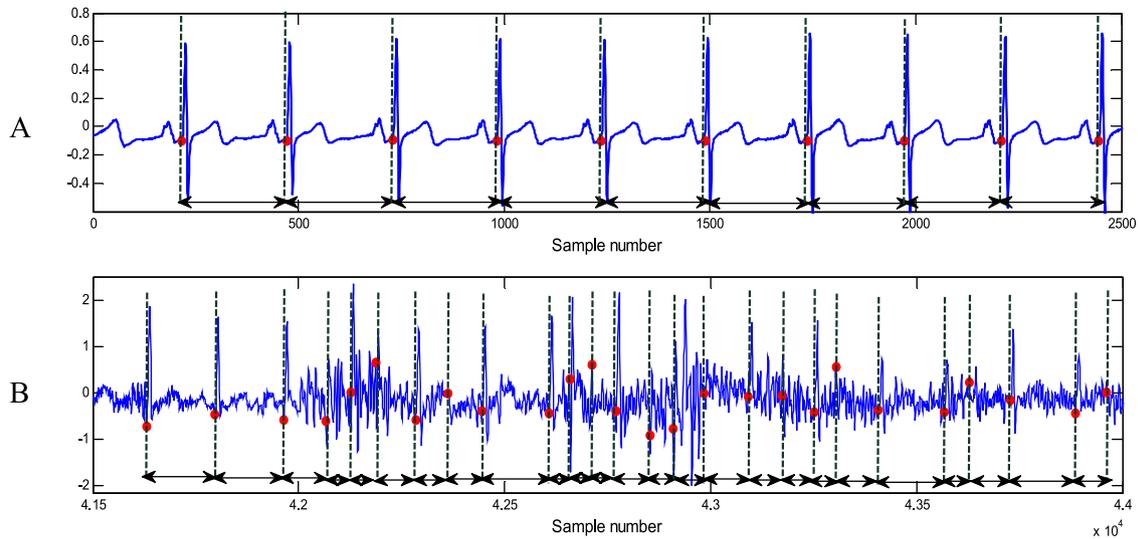

**Fig. 10.** Two ECG segments with same length (5 s) and different noise levels, (B) is more corrupted by noise and variance of the RR intervals in this segment is greater than the variance of the RR intervals in (A) which is completely clean.

der to find out if any FN and FP detections occurred in the distance between two consecutive windows where the algorithm switches from one signal to another. There are two situations on which occurrence of a FP or a FN is possible as described below:

1- The last detection of an ECG window is repeated in the next window of another signal as the first detection, when the algorithm switches to the detections of other signals in a window. In such situation it is possible that a FP occurs.
2- A heart beat located some milliseconds after a five second window of ECG ended, and in next window of detections of other signals, because of the delay correction mentioned earlier, heart beat transferred to the last detection window. In such situation when algorithm switches from ECG to other signals, a FN occurs.

In order to compensate these kinds of false detections, a search back strategy containing a FP remover and a FN remover function is employed in the last phase of the algorithm. These two functions compute the min and max permissive interval of two consecutive beats dynamically based on last 10 RR intervals as follow, Min RR interval = 0.5 (mean of last 10 RR intervals) and Max RR interval = 1.3 (mean of last 10 RR intervals). If RR interval of two consecutive beats is less than Min RR interval, FP remover looks for a FP detection among these two consecutive detections. Among these two beats, the one which the difference of its RR interval from mean of last 10 RR intervals is shorter is accepted as a TP and the other detection is considered as a FP and will be removed. If RR interval of two consecutive beats is greater than Max RR interval, FN remover looks for a FN detection among these two consecutive detections, and the local maximum in the distance of these two beats is detected and considered as a beat and would be added to the final detections.

### 3.8. Evaluation metrics

In order to score the proposed algorithm in this paper, created annotations by the algorithm on each of the database records are compared beat-by-beat to the reference annotations in which the exact location of beats are determined by some experts. A neighborhood of 300 ms centered at each reference beat is permitted and any beat detected in this neighborhood by the algorithm is accepted as a correctly detected beat. Four performance statistics are used in this paper for scoring and comparing this algorithm performance with the other available algorithms which are $Se_{average}$, $PPV_{average}$, Acc and $F_1$ measure [28], which gives an average of Se and PPV [29], and defined as:

$$Se_{average} = \frac{100}{n} \cdot \sum_{i=1}^{n} \frac{TP_i}{TP_i + FN_i} \quad (7)$$

$$PPV_{average} = \frac{100}{n} \cdot \sum_{i=1}^{n} \frac{TP_i}{TP_i + FP_i} \quad (8)$$

$$F_1 = 2 \cdot \frac{PPV \cdot Se}{PPV + Se} = \frac{2 \cdot TP}{2 \cdot TP + FN + FP} \quad (9)$$

$$Acc = \frac{TP}{TP + FN + FP} \quad (10)$$

where TP (true positive) denotes correctly detected beats or number of detected beats by proposed algorithm which are at the allowed neighborhood of the reference beats, FP (false positive) denotes number of detected beats which are not at the allowed neighborhood of any reference beat or more than one beat detected in allowed neighborhood of one reference beat and FN (false negative) denotes not detected beats or number of reference beats which there is no corresponding beats with them in detected beats of the proposed algorithm.

Some records in the database were shorter than 10 min, for example record 2041 has 20 s length. Although no matter how long the record is and the algorithm could work on it, there is a problem about scoring statistics. In calculation of $Se_{average}$ and $PPV_{average}$, all records regardless of their length have a same weight in averaging. It seems to be better to remove records whose lengths are shorter than 10 min from the database, for this kind of scoring.

### 4. Results

Fig. 11 shows different steps of ECG signal preprocessing, explained in Section 3.2.

In Fig. 12, the process of delay correction is depicted, explained in Section 3.4.

The results obtained by different ECG-based beat detectors, which are used in the presented technique on the dataset explained in Section 2.1, are assembled in Table 2, while the obtained results of different beat detection methods, which work based on multimodal data, are shown in Table 3. From Table 2 it could be seen that gqrs ECG beat detector obtained the best results among all ECG-based detectors, comparing this to the results presented in Table 3, it could be seen that approximately all of the beat detection methods which work based on multimodal data gathered better results than what gathered by gqrs. It could be concluded that by not sufficing on ECG signal and using the latent information about heart beats in other pulsatile signals, a better heart beat location estimation could be acquired and from the results reported in the Table 3 it could be concluded that the method presented in this paper gained better results in comparison with all other methods which are named in Table 3, and this method increased the sensitivity and positive predictivity of detection respectively about 3% and about 5% in comparison to the gqrs detector.

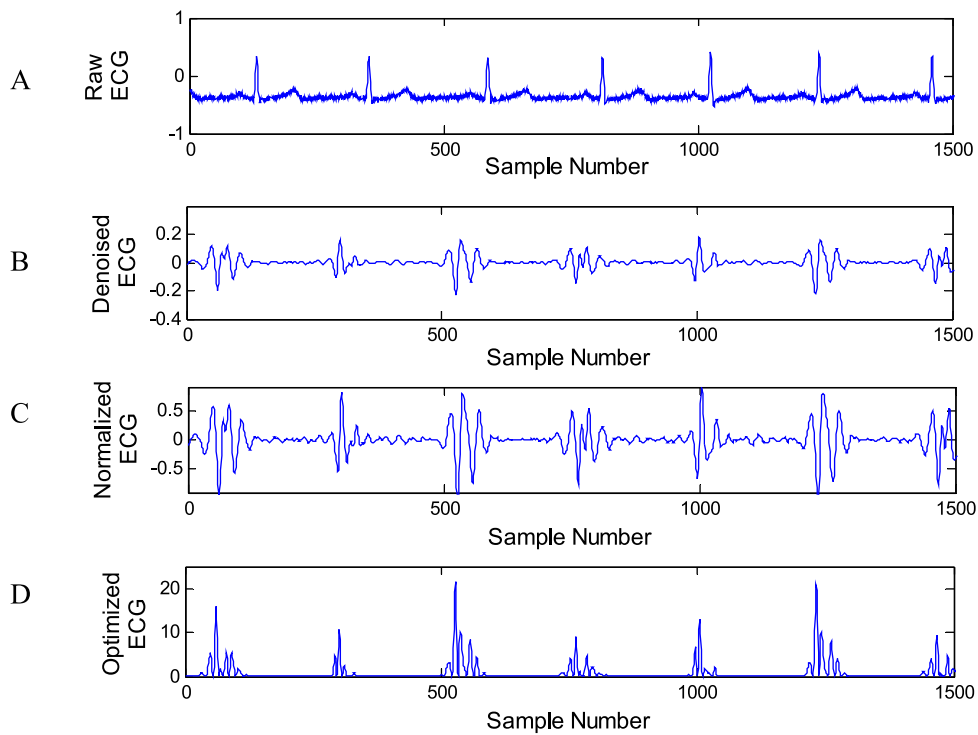

**Fig. 11.** Different steps of ECG signal preprocessing, (A) raw ECG signal, (B) summation of wavelet coefficients of d2, d3, and d4 levels of ECG signal wavelet decomposition, (C) normalized signal (acquired by Eq. (1)), and (D) optimized signal (acquired by Eq. (2)).

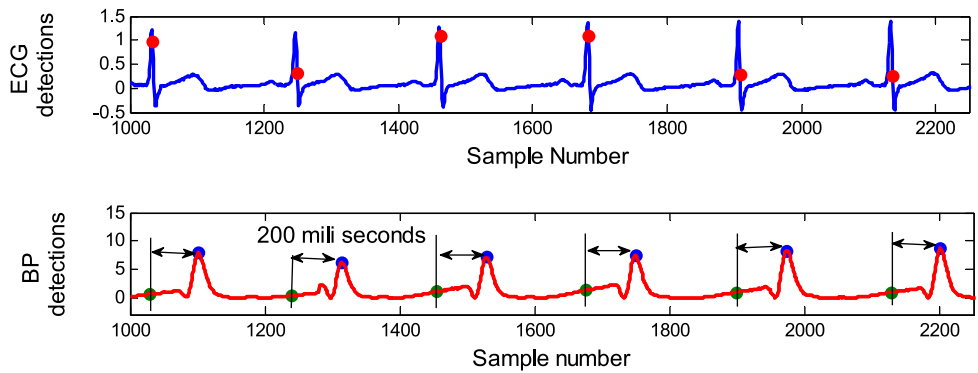

**Fig. 12.** Delay correction of BP detections, red bullets on ECG signal are detected beats on ECG and blue bullets on BP signal are detected beats on BP, dotted lines that cross the BP signal are drawn up along QRS complexes (beats location) of ECG signal, green bullets on BP signal are the location of BP beats after delay correction. (For interpretation of the references to colour in this figure legend, the reader is referred to the web version of this article.)

Table 2
Results gained by different single signal based detectors on database which is used in this study.

| Method name | Average Se (%) | Average PPV (%) | Average Acc (%) | Average F1 (%) |
| --- | --- | --- | --- | --- |
| Pan and Tompkins | 71.49 | 81.82 | 67.19 | 71.95 |
| Adaptive threshold | 83.31 | 59.71 | 71.57 | 75.82 |
| Absolute maximum | 80.81 | 81.78 | 75.04 | 80.81 |
| Local maximum | 89.93 | 79.4 | 74.67 | 81.03 |
| Gqrs | 92.51 | 91.17 | 84.90 | 91.49 |
| Wabp | 92.00 | 90.22 | 83.66 | 90.52 |

**Table 3**
Results gained by different algorithms which use multimodal recordings for robust heart beat detection on database which is used in this study.

| Author | Average Se (%) | Average PPV (%) | Average Acc (%) | Average F1 (%) |
| --- | --- | --- | --- | --- |
| Ghafari [4] | 94.09 | 92.96 | 89.99 | 93.26 |
| Thomas De Cooman [12] | 94.20 | 92.45 | 89.66 | 92.73 |
| Teo Soo King (unpublished) | 92.27 | 91.39 | 88.25 | 91.54 |
| Gierałtowski [18] | 93.70 | 93.66 | 90.88 | 93.42 |
| Sachi Vernekar (unpublished) | 95.1 | 94.6 | – | 94.5 |
| Johnson [30] | 94.1 | 94.2 | – | 93.7 |
| Marcus Vollmer [31] | 92.6 | 94.3 | – | 93.7 |
| Alistair E W Johnson [13] | 96.5 | 95.1 | – | 95.5 |
| This work | 95.47 | 96.03 | 93.11 | 95.62 |

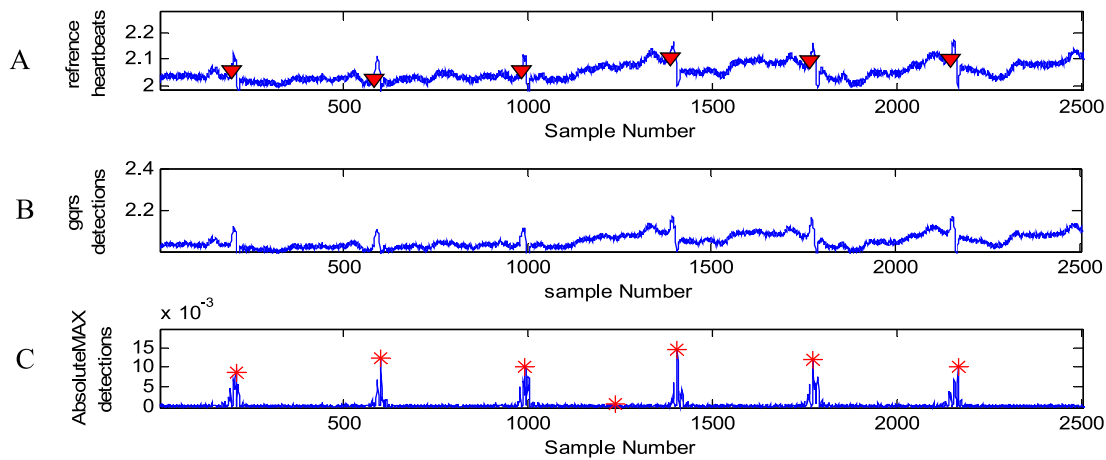

**Fig. 13.** Different peak detectors applied to the same ECG segment, (A) reference beats based on experts ideas, (B) detected beats by gqrs detector (no beats are detected because the amplitude of beats in this record are low), (C) detected beats by Absolute Max global detector.

The presented algorithm is associated with the 2014 Physionet challenge [30]. In this work, three general beat detectors are used in addition to the proprietary detectors for ECG and BP that makes this algorithm extendable for other kinds of pulsatile signals such as PPG. Another advantage of employing several peak detectors in this method can be characterized in odd signal shapes where proprietary detectors are not capable of detecting beats, in some of such signals general detectors are capable of detecting the exact location of beats correctly. The record 1683 is a good example of such cases; amplitude of the QRS complexes in this record is low, and none of the proprietary detectors could detect the location of peaks on ECG signal of this record, but as you can see in Fig. 13, Absolute Max general detector could detect heart beats correctly.

In this algorithm, paced beats are accounted and a strategy is employed for recognition of ECG signals containing paced beats; in first signal quality assessment step, such signals are recognized by a beat-by-beat comparison among detected beats of gqrs and wabp, such ECG signals are removed from database and location of beats in records which contain ECG with paced beats were extracted from BP or ART signals (Fig. 14).

In Table 2, the results gained by ECG-based detectors, which all use only ECG signal for heart beat location estimation, on all the 200 records presented in the database are shown; in addition to that, the results gained by wabp detector, which is also a single signal based detector that works on the hemodynamic signal with the lowest index available in each record, is shown in this table. Among all single signal beat detectors, the highest results belong to gqrs algorithm and after that wabp algorithm and among all general detectors, local max gained better results than other two detectors working on ECG.

In Table 3, the results of 8 best performing algorithms contributing in 2014 Physionet challenge with the presented algorithm in this paper are shown. These algorithms all used more than one signal of the records for heart beat detection and overall performance of them is better than any single signal based beat detector. As you can see in Table 3, highest F1 measure, 95.62%, belongs to the presented algorithm in this paper and it is about 0.12% higher than F1 measure of Alistair E W Johnson's algorithm [13] which gained better results among all other algorithms. Presented algorithm also performed better than Johnson's algorithm about 0.9% in PPV, but in Se this algorithm's performance is 1% better.

## 5. Discussion

Overall, the idea of increasing the robustness of beat detection using multimodal signals for estimation of exact location of heart beats was shown to be beneficial, especially when the ECG signal is corrupted by noise or not recorded correctly. In addition, in cases that diagnosing arrhythmia is the main goal of monitoring a patient, such approaches could be used because arrhythmia in some cases could be distinguished, in addition to ECG signal, on other signals which are related to the activity of heart especially pressure signals, such as BP. As an example of this case is the atrial fibrillation that causes quivering or irregular heartbeats and its effects appear on ECG and pressure signals. There are many studies about distinguishing normal beats and arrhythmia in a recording [32–34], but the algorithm presented in this paper is not going to find arrhythmia; the algorithm is based on the Physionet challenge 2014 and only aimed to classify detected beats in 2 classes, beats or not beats.

The algorithm presented here has some benefits over similar algorithms. Having so many detectors that are designed based on different approaches makes this algorithm robust against changes in signals shapes, for example when signals have reverse peaks or the amplitude of the peaks are very low, in such situations peaks could be detected by the use of these detectors. Another benefit of this work to similar works is having three scoring criteria in the fusion phase. There are some situations in which false detections could fulfil expectations of one of these three criteria but the two others recognize these false detections. For example, when all of the detections on a segment are FPs but the detections ratio fulfil the expectations of logical physiological range, in such cases regularity of detections and compatibility of detections could be useful for finding the best set of detections on other signals. Another benefit of this method to similar works is having a strategy for distinguishing noisy segments of ECG from clean ones with an adaptive method whose thresholds are calculated for ECG record proprietary. This algorithm also have some limitations; this algorithm is trained and tested on a database collected from people in resting situation, and it might do not have satisfying results on databases which are collected from people in exercising situation. Another limitation of this algorithm is in the records on which there is only one other signal which is related to the heart activity but ECG available in a record and this signal is too oscillatory because of noise. If detector detects lots of false detections on this signal but the number of detections be logical, algorithm might cannot recognize which signal is better for heart beat detection.

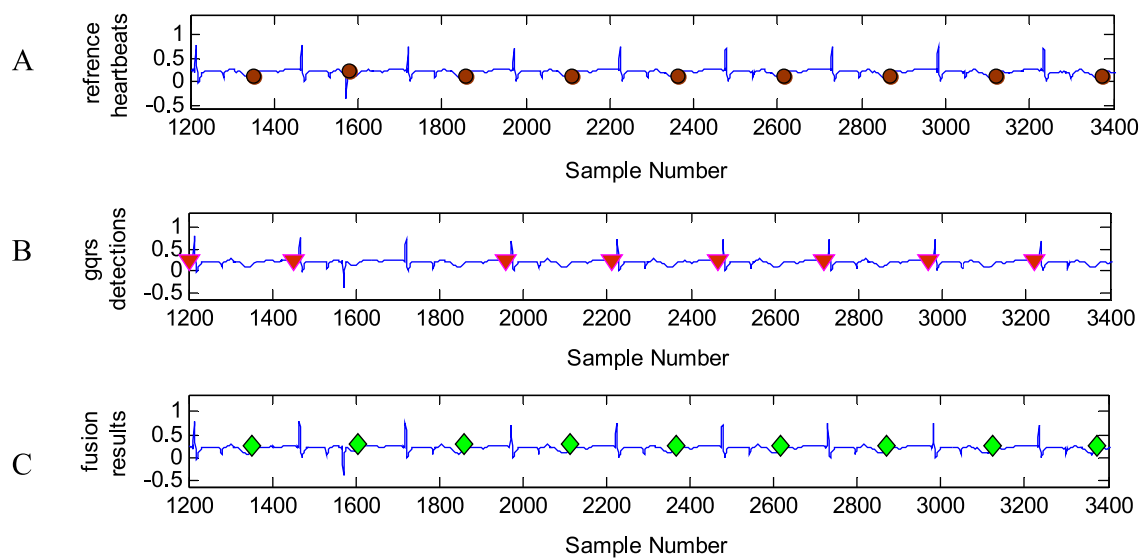

**Fig. 14.** Detected beats by the proposed algorithm on a recording in which ECG signal is containing paced beats. (A) Reference beats based on experts ideas, (B) detected beats by gqrs detector, (C) final beats based on fusion phase.

## 6. Conclusions

In this paper, a robust heart beat detection method with an accurate fusion strategy for fusing detected beats of different signals is presented. The fusion strategy employed in this paper was based on estimation of signals quality by different criteria. Estimated quality reflects the level of trustworthiness of different signals in each segment. This algorithm is designed in such a way that works on ECG, hemodynamic signals and SV but it could be extended to work on other pulsatile signals such as PPG. Above all, the presented method employed a strategy for recognizing if ECG is corrupted by noise or it is clean, and it could be very helpful for records that contain only one other signal but ECG which is suitable for heart beat detection.

## Conflict of interest

There is no conflict of interest.